\documentclass[preprint,12pt]{elsarticle}

\usepackage{amsmath,amssymb,amsfonts}
\usepackage{graphicx}
\usepackage{booktabs}
\usepackage{multirow}
\usepackage{array}
\usepackage{makecell}
\usepackage[hyphens]{url}
\usepackage[hidelinks]{hyperref}
\newcolumntype{C}[1]{>{\centering\arraybackslash}p{#1}}
\newcolumntype{R}[1]{>{\raggedleft\arraybackslash}p{#1}}

\journal{ISOFIC 2024}

\begin{document}

\begin{frontmatter}
\title{An Efficient Intrusion Detection System for Safeguarding Radiation Detection Systems}
\author[a]{Nathanael Coolidge}
\author[a]{Jaime González Sanz}
\author[a]{Li Yang}
\author[a]{Khalil El-Khatib}
\author[a]{Glenn Harvel}
\author[b]{Nelson Agbemava}
\author[c]{I Putu Susila}
\author[d]{Mehmet Yavuz Yağci}
\address[a]{Ontario Tech University, Canada}
\address[b]{Nuclear Regulatory Authority, Ghana}
\address[c]{National Research and Innovation Agency of Indonesia, Indonesia}
\address[d]{İstanbul University-Cerrahpaşa, Turkey}
\begin{abstract}
\footnote{
Preprint author original pre review. Accepted and Presented at ISOFIC 2024. The official proceedings version is available on the conference site.}Radiation Detection Systems (RDSs) are used to measure and detect abnormal levels of radioactive material in the environment. These systems are used in many applications to mitigate threats posed by high levels of radioactive material. However, these systems lack protection against malicious external attacks to modify the data. The novelty of applying Intrusion Detection Systems (IDS) in RDSs is a crucial element in safeguarding these critical infrastructures. While IDSs are widely used in networking environments to safeguard against various attacks, their application in RDSs is novel. A common attack on RDSs is Denial of Service (DoS), where the attacker aims to overwhelm the system, causing malfunctioning RDSs. This paper proposes an efficient Machine Learning (ML)-based IDS to detect anomalies in radiation data, focusing on DoS attacks. This work explores the use of sampling methods to create a simulated DoS attack based on a real radiation dataset, followed by an evaluation of various ML algorithms, including Random Forest, Support Vector Machine (SVM), logistic regression, and Light Gradient-Boosting Machine (LightGBM), to detect DoS attacks on RDSs. LightGBM is emphasized for its superior accuracy and low computational resource consumption, making it particularly suitable for real-time intrusion detection. Additionally, model optimization and TinyML techniques, including feature selection, parallel execution, and random search methods, are used to improve the efficiency of the proposed IDS. Finally, an optimized and efficient LightGBM-based IDS is developed to achieve accurate intrusion detection for RDSs.
\end{abstract}
\end{frontmatter}
\section{Introduction}

Over the last few years, governments worldwide have been actively exploring solutions to counter potential threats involving nuclear or other radioactive materials. Leading international organizations such as the International Atomic Energy Agency (IAEA) are pioneering the development of Radiation Detection Systems (RDSs) that can help with the early detection of threats from nuclear and other radioactive material \cite{r1}. These radiation detection systems are deployed in various applications, from monitoring environmental background radiation to enhancing security in public spaces such as stadiums and aircraft, and even in medical treatment settings \cite{r2} \cite{r3}. These radiation detection systems are deployed in various applications, from monitoring environmental background radiation to enhancing security in public spaces such as stadiums, ports, border crossings, and even in medical treatment settings \cite{r2} \cite{r3}. However, when malfunctioning, RDSs compromised by cyber-attacks may result in the need for rework, causing delays in diagnosis and treatment, which could prove lethal for some cancer patients.

It is crucial to highlight the vulnerabilities of these systems in real-world scenarios. For example, in nuclear facilities, the absence of robust IDS could allow attackers to manipulate data, posing significant risks to public safety. Similar challenges can occur in medical devices such as CT scans and radiation therapy machines, where the manipulation of data could delay vital treatment. A malfunctioning RDS compromised by cyber-attacks may fail to detect dangerous radiation levels, which can have severe consequences. For instance, in the context of environmental background monitoring, malfunctioning systems could lead people to be exposed to unhealthy levels of radiation, leading to an increased risk of cancer and other illnesses \cite{r2}. In medical scenarios like Computed Tomography (CT) scans, malfunctioning systems can cause distorted imaging, leading to potential misdiagnoses \cite{r3}. In nuclear facilities, malfunctioning RDSs could pose severe public safety risks. Similarly, accurate radiation monitoring is essential during large public events to mitigate the risk of radiation exposure and nuclear-based terrorist attacks.

Given the severe implications of compromised RDSs, these systems become prime targets for cyber-attacks. Cyber-attacks on digital infrastructure, initiated by threat actors, aim to either steal sensitive information or compromise the systems to hinder their functionality for malicious purposes \cite{r4}. One common method is through a denial of service attack (DoS), where the system is overwhelmed by a sudden burst of data packets designed to mimic legitimate traffic, thus disrupting normal operations and making detection challenging \cite{r5}. DoS (or the distributed version, DDoS) attacks are a particularly effective vector against RDS due to their ability to overwhelm systems with limited computational power. Despite their low cost and minimal resource requirements from the attacker's side, DoS attacks can cause significant downtime to the target \cite{r5}. DoS attacks, especially in cloud-connected RDS systems, can disrupt real-time radiation monitoring, delaying threat detection and response. Even if the RDS is not directly internet-connected, the servers storing the data in the cloud relaying the information to the user could be compromised, resulting in a DoS attack. This makes DoS attacks a critical threat vector for RDS.

To mitigate the risk of DoS attacks in RDSs, it is crucial to develop Intrusion Detection Systems (IDSs) that can monitor network traffic and detect abnormal DoS attack patterns. Machine Learning (ML) and data analytics models are effective methods for DoS attack detection by analyzing radiation datasets. This paper proposes a TinyML-based IDS framework that focuses on DoS attack detection for RDSs. TinyML presents a transformative approach to intrusion detection on low-resource devices such as mobile radiation detectors \cite{r6}. Deploying optimized ML models on TinyML-enabled devices can ensure real-time intrusion detection while minimizing power consumption, further enhancing the security of RDS in field applications.

To explore the impact of DoS attacks on RDS, the proposed framework first utilizes K-Means clustering to identify anomalies within a real radiation dataset, followed by the application of the Synthetic Minority Over-sampling Technique (SMOTE) to amplify these anomalies and create a synthetic training and testing dataset for RDS. Subsequently, we assess the performance of several supervised ML models, including Random Forest (RF), SVM, logistic regression, and LightGBM, to determine the most effective model for DoS attack detection. The choice of these ML models in the proposed framework is guided by their proven efficiency in anomaly detection tasks and their ability to handle high-dimensional data. LightGBM, in particular, was selected after evaluation for its low computational requirements and fast training speed, making it a suitable candidate for deployment on resource-constrained RDS devices. Additionally, TinyML techniques for LightGBM, including feature selection, parallel execution, and random search-based hyperparameter optimization, are used to optimize the performance and efficiency of LightGBM to further improve model effectiveness. The proposed IDS framework is evaluated on a real radiation dataset from Safecast \footnote{Safecast API datasets: https://github.com/Safecast/safecastapi/wiki/Data-Sets}. 

This paper makes the following contributions:
\begin{enumerate}
\item A robust IDS for RDSs using LightGBM to safeguard against cyber threats such as DoS attacks while ensuring the integrity and accuracy of radiation readings;
\item A novel attack data generation method that applies K-Means clustering and SMOTE to effectively handle class imbalance by generating synthetic anomalies, simulating potential DoS attacks on RDS data;
\item Evaluation of multiple machine learning models, including LightGBM, Random Forest, logistic regression, and SVM, to identify the most effective model for detecting malicious data in RDS.
\item A TinyML-based LightGBM model that integrates multiple TinyML techniques (i.e., feature selection, parallel execution, and hyperparameter optimization) to optimize model performance and efficiency, enabling real-time, low-latency, and energy-efficient intrusion detection.
\end{enumerate}

The rest of the paper is organized as follows: Section 2 presents related work, including the generation of synthetic data using SMOTE and the development of effective ML algorithms for DoS detection. Section 3 describes the methodology used to select suitable features from the radiation dataset for ML analysis. It also outlines effective data preprocessing, hyperparameter optimization, and TinyML, procedures for ML model performance improvement. Section 4 presents and discusses the experimental results. Section 5 concludes the paper.

\section{Related Work}

\subsection{Data Balancing and Synthesis Data Generation}

An ongoing challenge with training machine learning models is class imbalance. In most datasets, there is an imbalanced ratio of true positive results and true negative results in the target column. The basic way to overcome this is by either undersampling, which removes data from the majority class, or oversampling, which increases the samples in the minority class through duplication or artificial generation. Among these techniques, oversampling often achieves better performance than undersampling, as no data is lost \cite{r7}. However, because random oversampling methods simply randomly replicate data in the minority class, this causes the model to suffer from overfitting \cite{r7}. Chawla \textit{et al.} \cite{r8} introduced the Synthetic Minority Oversampling Technique (SMOTE) to address the issue of overfitting. SMOTE improves performance by integrating K-nearest neighbors (KNN) to generate synthetic data based on slightly tweaking similar features. This process creates a more accurate representation of the minority class. 

While SMOTE helps to handle class imbalance, using SMOTE alone creates issues such as noise samples, overlapping samples, and boundary samples \cite{r9}. To mitigate these issues, many previous studies have proposed the use of K-Means in combination with SMOTE. K-Means is a clustering technique to filter out the minority class and cluster similar features amongst samples. Xu \textit{et al.} \cite{r7} explored the use of KN-SMOTE, which is a combination of K-Means and SMOTE, in the context of the C4.5 decision tree algorithm, which is widely used in the medical field but struggles with imbalanced data. Xu \textit{et al}. proposed a method that involves using K-Means to cluster all the samples, then filtering the results to gather the “safe samples”, which are the samples that have not changed classes after K-means clustered them, and finally using SMOTE to replicate these “safe samples”. In another study, Liu \textit{et al.} \cite{r10} explored the use of KN-SMOTE, which is called KM-SMOTE in the work, on Rockburst data. Rockburst is a violent expulsion of rock from the walls of a mine, and predicting its intensity is crucial for safety. However, the data is often imbalanced, meaning one class (e.g., high intensity) is significantly less prevalent than the other(s) (e.g., low intensity). This imbalance can lead to biased models that favor the majority class \cite{r10}. 

In conclusion, SMOTE addresses class imbalance by generating synthetic minority samples, but combining it with K-Means clustering improves sample quality by reducing noise and preventing overfitting. This makes the K-Means and SMOTE combination particularly effective for attack data generation in anomaly detection of RDSs, where minority class data is crucial. While prior work has focused on intrusion detection in traditional network environments, this paper addresses the gap in specialized IDS for RDS. By combining SMOTE with K-Means clustering, we create a more robust synthetic dataset to simulate DoS attacks, improving model performance compared to previous methods.

\subsection{Machine Learning for DoS Detection}

ML models have proven effective in detecting DoS attacks by classifying network traffic, with LightGBM and Random Forest performing particularly well. 

Bakhareva \textit{et al.} \cite{r11} explored the use of ML methodologies to identify and classify malicious traffic on a network. By using what, at the time, was the latest dataset and measuring performance, they tested the CatBoost and LightGBM ML algorithms. Binary classification (attack or benign) and classification of attacks (Dos/DDoS, Bruteforce, Bot, Infiltration, Portscan, etc.) were tested with both algorithms. It was concluded that both algorithms work well for the binary classification of network traffic and as well as multi-class classification of threats \cite{r11}. It is also pointed out that these algorithms can be used to identify new types of attacks. Dener \textit{et al.} \cite{r12} proposed a classification-based DoS intrusion detection system. The proposed approach combines SMOTE with Tomek-Link for data balancing with LightGBM as a ML algorithm. It was demonstrated that this approach was effective in big data environments, and that it was more successful than previous approaches, such as using the logistic regression or Random Forest algorithms \cite{r12}. Rani \textit{et al.} \cite{r13} evaluated different ML algorithms when detecting DDoS attacks in Device-to-Device (D2D) communications. These communications happen in close proximity while being dynamic, and the devices have low processing capabilities. These devices also have a limited battery life, so it was important to ensure the accurate detection of harmful data, while using minimal resources. The work demonstrated that different ML classification techniques, such as LightGBM, Random Forest, XGBoost, and AdaBoost, accurately identified DDoS, SYN, and slow loris attacks. Furthermore, the implementation proposed demonstrated minimal CPU and memory usage, thereby conserving battery power \cite{r13}. Liu \textit{et al.} \cite{r14} implemented similar methodologies by which similar results were obtained. On a dataset with imbalanced data, SMOTE was applied and then several supervised ML models were compared. LightGBM also performed best, overperforming models such as Random Forest and SVM, among others.  Similarly, Liu \textit{et al.} \cite{r15} also implemented an intrusion detection system using oversampling techniques that later compared several supervised ML models. LightGBM also performed best. It is worth mentioning that the oversampling technique used was ADASYN, a variation of SMOTE. We consider this technique to be a potential improvement to our current method.

In summary, LightGBM was found to be particularly suitable for battery-constrained environments due to its efficient handling of large datasets. Its use of histogram-based algorithms allows for faster computations with lower memory usage, making it an optimal choice for real-time intrusion detection in RDSs.

\section{Methodology}

\subsection{Data Generation}

In order to develop a ML model for the proposed IDS in RDSs, a dataset obtained from Safecast was analyzed. This dataset, known as the “Daily export”, contains radiation readings from “bGeiGie” radiation detection devices from all over the world, compiled in Safecast’s API and their GitHub page.

The radiation dataset includes both radiation readings and other factors grouped into a large dataset. These readings are only differentiated by a parameter named “unit”. By filtering this dataset to include only values measured in microsieverts per hour (“µSv/h”), which indicate radiation levels, a more manageable dataset that contains the key information can be obtained. 

While the data contains various parameters detailing when and where the data was recorded, the following features are considered to be key features for IDS development:
\begin{enumerate}
\item \textbf{Latitude and longitude}: while recorded in two columns, they can be used together to precisely identify where each reading was taken
\item \textbf{Value}: Represents the radiation measurement in microsieverts per hour after filtering the dataset to only include radiation readings.
\item \textbf{Device ID}: A unique identifier for the device that captured the radiation reading.
\item \textbf{Uploaded time}: Expressed in minutes, seconds, and tenths, this value indicates when the data was uploaded. This time is more precise than the time in which the data was captured.
\end{enumerate}

The existing RDS dataset is composed only of normal, unaltered data. To create an IDS capable of detecting threats, it is necessary to create synthetic data that simulates data logs that are similar to those a threat actor would send. We achieve this by leveraging anomalies in the dataset and creating the aforementioned replicated attack data based on those. These anomalies are a small amount of data logs that numerically deviate from standard patterns.

To identify these anomalies, unsupervised anomaly detection methods were used. Specifically, K-Means clustering was used to analyze and identify patterns in the dataset, and group logs into different clusters that share similarities. By analyzing the resulting clusters, it is possible to identify clusters with anomalies. K-means is simple, yet efficient and works exceptionally well in large datasets with high dimensionality, such as radiation reading datasets \cite{r9}. 

In our experiment, two types of anomalous clusters were discovered. The first type contained radiation readings close to 0, as a simple DoS attack could manipulate the system to read zero values, making the background radiation appear weak and potentially causing the system to overlook dangerous levels of radiation. The second type comprised clusters with a few logs reporting extremely high radiation readings, which could indicate either sensor malfunctions or a targeted DoS attack aimed at generating false alarms. Once the anomalies have been identified, a temporal label was added to the dataset. This was done in order to differentiate the anomalies from normal data logs. This allows for the generation of synthetic data that simulates potential cyberattacks.

Synthetic Minority Over-sampling Technique (SMOTE) was applied to the anomalous data. This method is more efficient than the random over-sampling method that simply replicates the logs, as SMOTE creates higher-quality data with more variation and realism, making the resulting IDS more robust. By generating synthetic samples of DoS attack samples, SMOTE ensures that the model receives enough diverse samples to learn from, leading to a more robust IDS capable of detecting rare but critical cyber-attacks (i.e., DoS). After generating synthetic data, Gaussian noise was introduced to the generated DoS attack samples to prevent the machine learning models from overfitting to artificial patterns that may arise during the data generation process. By adding noise, we mimic real-world variations and uncertainties, enhancing the model’s ability to generalize to new, unseen threats.

\subsection{Proposed IDS Development}

With the finalized dataset—comprising both synthetic data and Gaussian noise—the data was split into training and testing sets. Various supervised ML models were trained on the dataset to evaluate which performed best. The models analyzed include Random Forest, logistic regression, Support Vector Machine (SVM), and Light Gradient-Boosting Machine (LightGBM).

Random Forest is an ensemble learning method that constructs multiple decision trees during training and outputs the mode of the classes (classification) or mean prediction (regression) of the individual trees \cite{r16}. It reduces overfitting by averaging multiple trees, which improves accuracy and robustness. Random Forest can handle large datasets with higher dimensionality, making it effective for detecting patterns in network traffic data. Its ability to handle imbalanced datasets and provide feature importance makes it a strong candidate for identifying DoS attacks. 

Logistic regression is a statistical model that uses a logistic function to model a binary dependent variable \cite{r17}. It estimates the probability that a given input belongs to a particular class (e.g., attack or normal traffic) based on the input features. Logistic regression is simple, interpretable, and efficient for binary classification tasks. It works well when the relationship between the features and the target variable is linear, making it suitable for detecting DoS attacks in network traffic data.

SVM is a supervised learning model that finds the optimal hyperplane that maximizes the margin between different classes in the feature space \cite{r18}. It can handle both linear and non-linear classification by using kernel functions. SVM is effective in high-dimensional spaces and can model complex decision boundaries, making it suitable for detecting DoS attacks. Its ability to handle non-linear relationships and robustness to overfitting are advantageous for network traffic analysis. 

LightGBM is a gradient-boosting framework that uses tree-based learning algorithms. It builds trees leaf-wise rather than level-wise, which can result in faster training and better accuracy. LightGBM is highly efficient and scalable, making it suitable for large datasets with high-dimensional features. Its ability to efficiently handle imbalanced data makes it a strong candidate for detecting DoS attacks in network traffic data \cite{r19}.

Due to the high efficiency and scalability of LightGBM, and based on the findings in the literature review section, it has a high potential for being the best-performing model.

\subsection{Hyperparameter Optimization and TinyML Implementation}

After selecting the best-performing model, hyperparameter optimization was implemented to fine-tune the model to maximize its performance on the radiation dataset. Hyperparameter optimization is the process of automatically tuning the hyperparameters of ML models to design optimized ML models. For example, tuning the number of decision trees and the tree depth in models such as random forest and LightGBM models is essential for enhancing detection accuracy and minimizing false positives.

Random search was selected in the proposed IDS for hyperparameter optimization. Random search is an efficient and scalable method that explores the hyperparameter space by randomly selecting hyperparameter combinations and evaluating their performance. This method significantly reduces computation time compared to an exhaustive search (i.e., grid search) while yielding improved results. 

Additionally, due to its low computational complexity, LightGBM is ideal for resource-constrained environments such as TinyML implementations of the model. TinyML allows ML models to run in an extremely compact size, making it ideal for small mobile radiation detection systems \cite{r6}. Optimizing memory usage and processing power is critical in TinyML to ensure efficient, real-time detection of DoS attacks.

To adapt LightGBM for TinyML, several key steps were taken \cite{r6}:
\begin{enumerate}
\item \textbf{Feature Selection}: The first step in adapting LightGBM for TinyML was to reduce model complexity by selecting the most important features. We used feature importance metrics to identify and retain only the features that contributed to 90\% of the model’s predictive power. This significantly reduced memory usage and processing time, which is essential for devices with limited resources.
\item \textbf{Parallel Execution}: To maximize efficiency in resource-constrained environments, we leveraged LightGBM’s built-in support for parallel execution. This allows the training process to be distributed across multiple CPU cores, speeding up both training and inference. In TinyML applications, where devices may have multi-core processors, parallel execution improves the model’s ability to process data in real time, crucial for detecting DoS attacks.
\item \textbf{Hyperparameter Tuning for Efficiency}: During the random search-based hyperparameter optimization process, we specifically focused on fine-tuning parameters such as n\_estimators and tree depth (max\_depth). By setting the number of trees to a lower range, we were able to minimize the computational cost of each prediction while maintaining reasonable accuracy. Lowering tree depth and limiting the number of leaves also contributed to reducing memory usage and ensuring the model could efficiently run on resource-constrained devices.
\end{enumerate}

By implementing feature selection, parallel execution, and hyperparameter tuning, LightGBM was successfully adapted for TinyML, enabling real-time intrusion detection in resource-constrained environments. These optimizations allowed the IDS to function efficiently in mobile radiation detection systems and other resource-constrained applications, providing real-time detection of DoS attacks without overburdening the device's hardware resources.

\section{Performance Evaluation}

The proposed IDS was then built using a filtered version of the dataset, that included only the radiation readings and important factors mentioned in Section 3. To generate the synthetic data that represents DoS attacks, the following methodology was used:

An initial preprocessing for the clustering process was conducted. The following parameters were retained to identify anomalies in the data:
\begin{enumerate}
    \item Captured Time
    \item Latitude
    \item Longitude
    \item Value (µSv)
    \item Uploaded Time
\end{enumerate}

When analyzing the recorded times, since they were originally represented in the format [“yyyy-mm-dd hh:mm:ss.ms”], they were transformed to Unix timestamps instead. This represents the number of seconds that have passed since the Unix epoch (00:00:00 Thursday, January 1, 1970)\textbf{.} Afterward, min-max scaling was applied. 

We then performed clustering using K-Means to identify anomalies in the data. To determine an optimal number of clusters, we performed a random search to evaluate values ranging from 5 to 100. We found that 65 clusters were the minimum number of clusters that could clearly and consistently identify anomalies in the data that show a very low or very high radiation reading. Although more clusters could also identify the anomalies, they require additional computational power and processing time.

Readings considered very low either had a flat value of 0—which is anomalous due to the presence of background radiation—or values extremely close to 0 (less than half the next smallest average reading). The very high readings were two values that recorded a reading of 44 and 66 µSv, respectively, while the third highest value was 0.3 µSv.

After marking the anomalies on the original dataset, we performed a second round of preprocessing. The process was quite similar to the one used for clustering, but the device IDs were included by applying one hot encoding. This creates a new column for each device ID. If the ID matches, it is represented as a 1, otherwise it is a 0. The number of different device IDs in our dataset was relatively small, so it was implemented with ease. 

Next, 40,000 new readings were created using smote based on the anomaly data, which was labeled in the previous step. The original unaltered data had 130,000 data samples recorded. This new set of readings represents our DoS attack. Finally, Gaussian noise is applied to the whole data set. Then, the dataset is ready for supervised ML.

Each of the four models described in Section 3.2 (i.e., LightGBM, Random Forest, logistic regression, and SVM) was trained using the same partition of the dataset. They were fitted without any modifications or hyperparameter optimization. Then, each model predicted on the same “test data''. Each model's performance was evaluated using the following metrics: accuracy, precision, recall, F1-score, and prediction time.\textbf{ }These metrics provide a comprehensive view of the model's effectiveness in balancing accuracy, recall (detecting real threats), and precision (minimizing false alarms), ensuring efficient and accurate detection in real-time applications.

The performance of the models — LightGBM, Random Forest, logistic regression, and SVM — can be found in Table 1. Among these models, LightGBM performed best, achieving the highest accuracy at 98.174\% and the highest F1-score of 96.186\%. LightGBM also has the second-best recall (98.421\%), and second-best prediction time per sample (0.725 µs). These results demonstrate its ability to balance the detection of true threats while minimizing false alarms, which is crucial for effective intrusion detection.

\begin{table}[t]
\centering
\caption{Comparison of ML Models}
\label{tab:1}
\renewcommand{\arraystretch}{1.2}
\scalebox{0.75}{
\begin{tabular}{|>{\centering\arraybackslash}m{7em}|>{\centering\arraybackslash}m{5em}|>{\centering\arraybackslash}m{5em}|>{\centering\arraybackslash}m{5em}|>{\centering\arraybackslash}m{5em}|>{\centering\arraybackslash}m{7em}|}
\hline
 \textbf{Model} & \textbf{Accuracy (\%)} & \textbf{Precision (\%)} & \textbf{Recall (\%)} & \textbf{F1-score (\%)} & \textbf{Prediction Time Per Sample (µs)} \\ \hline
 LightGBM & 98.174 & 94.049 & 98.421 & 96.186 & 0.725 \\ \hline
 Random Forest & 98.070 & 94.717 & 97.166 & 95.926 & 1.533 \\ \hline
 Logistic Regression & 97.628 & 93.564 & 96.495 & 95.007 & 0.058 \\ \hline
 SVM & 98.076 & 92.397 & 100.000 & 96.048 & 312.698 \\ \hline
\end{tabular}}
\end{table}

Although SVM achieved a perfect recall of 100\%, this result is likely due to overfitting, as indicated by its much lower accuracy (98.076\%) and an extremely high prediction time per sample of 312.698 µs. While SVM successfully identified all the malicious data, its impractical prediction time and potential overfitting make it unsuitable for real-time IDS applications. This result underscores the importance of balancing recall with prediction time, as overfitting can lead to a model that, while thorough, is inefficient and difficult to deploy in real-world environments. 

Logistic regression, though having the fastest prediction time per sample at 0.058 µs, exhibited lower accuracy (97.628\%) and F1-score (95.007\%). While fast, its lower performance in terms of accuracy suggests it may not be suitable for more complex or imbalanced datasets, where misclassifying malicious data could lead to serious consequences.

Random Forest, with an accuracy of 98.070\% and an F1-score of 95.926\%, also performed well but was outshined by LightGBM in terms of both prediction time and model accuracy. Random forest's ability to handle imbalanced data was evident in its high precision (94.717\%), but the marginally slower prediction time (1.533 ms) makes LightGBM a more attractive option for real-time applications.

In order to further improve the model, hyperparameter optimization was done with a randomized search. The hyperparameters explored consist of: “n\_estimators”, “max\_depth” and “num\_leaves”. The hyperparameter tuning results can be found in Table 2.

\begin{table}[t]
\centering
\caption{Hyperparameter Optimization Parameters}
\label{tab:2}
\renewcommand{\arraystretch}{1.2}
\scalebox{0.75}{
\begin{tabular}{|>{\centering\arraybackslash}m{10em}|>{\centering\arraybackslash}m{7em}|>{\centering\arraybackslash}m{7em}|}
\hline
 \textbf{Hyperparameter} & \textbf{Search Space} & \textbf{Best Value} \\ \hline
 n\_estimators & [10,50] & 42 \\ \hline
 max\_depth & [3,8] & 6 \\ \hline
 num\_leaves & [4,20] & 18 \\ \hline
\end{tabular}}
\end{table}

The results obtained from the model after the hyperparameter optimization, compared to those before it, are presented in Table 3. After hyperparameter optimization, LightGBM saw further improvements, with its accuracy rising to 98.247\%, recall increasing to 98.732\%, and F1-score improving to 96.343\%. Additionally, due to the use of TinyML techniques, the prediction time per sample decreased from 0.611 µs to 0.521 µs. The hyperparameter adjustments, particularly decreasing the number of estimators from 100 to 42 and setting the maximum depth to 6, allowed the model to identify patterns in the data more effectively and efficiently, leading to better generalization and lower resource consumption. Reducing the number of leaves from 31 to 18 helped prevent overfitting, ensuring that the model performed well on unseen data.

\begin{table}[t]
\centering
\caption{LightGBM Optimization Results}
\label{tab:3}
\renewcommand{\arraystretch}{1.2}
\scalebox{0.75}{
\begin{tabular}{|>{\centering\arraybackslash}m{7em}|>{\centering\arraybackslash}m{5em}|>{\centering\arraybackslash}m{5em}|>{\centering\arraybackslash}m{5em}|>{\centering\arraybackslash}m{5em}|>{\centering\arraybackslash}m{7em}|}
\hline
 \textbf{Model} & \textbf{Accuracy (\%)} & \textbf{Precision (\%)} & \textbf{Recall (\%)} & \textbf{F1-score (\%)} & \textbf{Prediction Time Per Sample (µs)} \\ \hline
 LightGBM & 98.174 & 94.049 & 98.421 & 96.186 & 0.725 \\ \hline
 TinyML Optimized LightGBM & 98.247 & 94.067 & 98.732 & 96.343 & 0.521 \\ \hline
\end{tabular}}
\end{table}

The superior accuracy and low resource consumption of TinyML-optimized LightGBM make it ideal for real-time applications, such as RDS, where immediate responses are critical. This ensures that potential threats are identified promptly without compromising system performance, demonstrating why LightGBM and TinyML are the optimal choices for real-time intrusion detection in RDS.

In summary, LightGBM emerged as the best model due to its balance of high accuracy, low prediction time, and superior handling of imbalanced data. These characteristics make it highly suitable for intrusion detection in radiation detection systems, where real-time performance and robustness are paramount. TinyML techniques further improve the efficiency of the proposed IDS to achieve real-time DoS attack detection.

\section{Conclusion}

Radiation Detection Systems (RDSs) play an important role in protecting public safety across various sectors, including environmental monitoring, healthcare, and nuclear facilities. Ensuring the integrity of these systems and the accuracy of the radiation readings they generate is crucial. To safeguard against cyber threats, such as tampering or data manipulation, the implementation of an Intrusion Detection System (IDS) is essential. An effective IDS must be capable of distinguishing between genuine radiation data and maliciously altered data. In this paper, we focused on simulating one type of attack—a Denial of Service (DoS) attack—to demonstrate how false data can be detected within RDS environments. To address class-imbalance in the dataset, we utilized a combination of K-Means clustering and SMOTE for oversampling. K-Means was effective in identifying significant anomalies, which were then replicated through SMOTE to create a balanced dataset. Our experiments demonstrated that the TinyML-optimized LightGBM, when applied to this enhanced dataset, successfully detected false data with a high degree of confidence. The results highlight the superiority of the proposed TinyML-optimized LightGBM, particularly for its efficiency in low-resource environments. For future work, we plan to explore more advanced oversampling techniques, such as ADASYN and DEAGO, to further improve the balance and quality of synthetic data. Additionally, deploying the IDS on resource-constrained devices using advanced TinyML techniques could establish a new standard for real-time intrusion detection in RDSs globally. These advancements will establish a new standard for securing RDS globally, ensuring both high detection accuracy and real-time performance in environments facing increasing nuclear threats. 

\section{Acknowledgment}

This work is partially funded by the Natural Sciences and Engineering Research Council of Canada (NSERC). The work is also supported by the on-going International Atomic Energy Agency (IAEA) Coordinated Research Project (CRP) J02017 entitled “Enhancing Computer Security for Radiation Detection Systems.”

\end{document}